\documentstyle[11pt]{article}
\begin{document}
\baselineskip 18pt
\begin{titlepage}
\centerline{\large\bf A Constraint on EHNS Parameters from
Solar Neutrino Problem }
\vspace{2cm}

\centerline{ Yong Liu, Jing-Ling Chen and Mo-Lin Ge }
\vspace{0.5cm}
{\small
\centerline{\bf Theoretical Physics Division}
\centerline{\bf Nankai Institute of Mathematics}
\centerline{\bf Nankai University, Tianjin 300071, P.R.China}
}
\vspace{2cm}

\centerline{\bf Abstract}

We suppose that the solar neutrino problem is only due to the mechanism
introduced by Ellis, Hagelin, Nanopoulos and Srednicki (EHNS), then
a lower limit constraint on EHNS parameters is obtained. We find that
$
\gamma\; \geq \; 7.4\times 10^{-22}\;GeV
$
if
$
\alpha < 2\; \gamma
$
and
$
\alpha\; \geq \; 1.5\times 10^{-21}\; GeV
$
if
$
\alpha > 2\; \gamma
$, this limit is consistent
with the upper limits extract from the $K^0-\overline{K^0}$
system and can be detected in the future experiments.\\
\\
PACS number(s): 14.60.Pq  03.65.Bz

\vspace{2cm}
\end{titlepage}

\centerline{\large\bf A Constraint
                      on EHNS Parameters from Solar Neutrino Problem }
\vspace{1cm}

\noindent {\bf I. Introduction.}

According to Hawking's postulate on the evaporation of black holes[1-2],
a pure quantum mechanical state may evolve into mixed states. Although
this violation of quantum mechanics could cause a lot of problem in physics,
such as a serious conflict between energy-momentum conservation
and locality[3] and
CPT violation[4-5] etc., the theory in this field has been set up.

To include the quantum mechanics violation owing to Hawking's quantum
gravity effect, in 1983, J.Ellis, J.S.Hagelin, D.V.Nanopoulos and
M.Srednick put forword a modified hamiltonian equation of motion for
density matrix and used it to interpret upper bounds on the violation
of quantum mechanics in different phenomenological situation[6]. Soon
after, T.Banks, L.Susskind and M.E.Peskin (BPS)[3] proposed a generic
form of the modified evolution equations for density matrix which
preserves the linearity, locality in time, and conservation of probability.
Recently, motivated by EHNS and BPS and the other researchers, B.Reznik[7]
proposed another modified Liouville equation which constitutes a linear,
local,and unitary extension and thus allows a quantum mechanical system
for unitary evolution between pure and mixed states. But different from
the exponentially decaying (or exponentially increasing) solution of EHNS
mechanism, this model yields a oscillatory modification. However, both
mechanisms can lead to CP violation. By analysing the experimental datum
of CP violation in $K^0-\overline{K^0}$ system, Ellis et al.[8-9] and
Peskin et al.[10] determined the values of two of the three EHNS parameters
independently. All the analysis gives the upper limits of EHNS parameters
approach the range $ 0(m_k^2/M_{pl})\approx 2\times 10^{-20}GeV$.
In the meantime, Ellis et al. determin these parameters by theoretical
consideration furthermore. Based on string theory[10-11] and display a
logarithmic divergence in the density matrix of a scalar field in the
presence of an Einstein-Yang-Mills black hole in four dimensions[12],
they get the values of the size of the estimate above. The latest estimate
of these parameters is[20]
$$
\alpha \leq 4\times 10^{-17} GeV ,\;\;\;\;
\mid \beta \mid \leq 3\times 10^{-19} GeV ,\;\;\;\;
\gamma \leq 7\times 10^{-21} GeV .
$$
But, if EHNS mechanism is a correct physical principle, the lower limits of
these parameters should be known when we do not know their exact values yet.

On the other hand, the solar neutrino problem which consists of the deficit
of observed neutrino emitted from the Sun with respect to the theoretically
expected amount has been with the physicists for more than 30 years. To
resolve this puzzle, a class of solutions such as neutrino oscillation and
the matter effect[13-15], neutrino magnetic moment[16], and a combination
of both etc. have been developed. In Ref.[19], we have found that the EHNS
mechanism can affect the neutrino oscillation behaviors greatly and hence
may be taken as a new solution of the solar neutrino problem[19].

In this work, we suppose that the solar neutrino deficit is due to EHNS
mechanism, so we can get a constraint on the lower limits of the EHNS
parameters. To ensure this paper is selfcontained, we first introduce
EHNS mechanism follwing Huet and Peskin[10] in section two and list the
relative formulas and datum about neutrino oscillation in section three,
then, in section four, we show how to estimate the lower limit constraint
on the EHNS parameters by using neutrino oscillation datum. The conclusion
and discussion are given in the final section.\\

\noindent{\bf II. EHNS Mechanism}

In conventional quantum mechanics, the evolution of density matrix obeys
Liouville equation:
\begin{equation}
i\frac{d}{dt}\rho=[H,\rho].
\end{equation}
This equation guarantees that the probability is conserved and the purity
of the state is not changed in the evolution of a system.

For a two states system, the density matrix can be expanded by using the
Pauli sigma matrix:
\begin{equation}
\rho=\rho^0 1+\rho^i \sigma^i
\end{equation}
where $i=1,2,3$ and repeat index represents summation.
It is evidently that $\rho^0=\frac{1}{2}$ and to ensure the density
of probability is semi-positive definite, $\rho^i$ should satisfy the
following constraint
\begin{equation}
(\rho^0)^2 \geq \sum_{i=1}^3 (\rho^i)^2 .
\end{equation}
When expanding the hamiltonian in the same way, Eq.(1) becames
\begin{equation}
\frac{d}{dt}\rho=2\epsilon^{ijk}H^i\rho^j\sigma^k
\end{equation}
where $\epsilon^{ijk}$ is the totally antisymmetric tensor and
\begin{equation}
\epsilon^{123}=+1
\end{equation}
For a non-Hermitian Hamiltonian, the case is little more complicated,
we refer the readers to the original paper of Huet and Peskin[10] or
Ellis et al[8][20].

Now, to include the modification due to Hawking's quantum gravity
effect which permits quantum mechanical system evolve into
mixed states, the most general linear terms
\begin{equation}
-h^{0j}\rho^j 1-h^{j0}\sigma^j -h^{ij}\sigma^i \rho^j
\end{equation}
should be added to Eq.(4). But there are two natural restrictions on
these terms: probability conservation and entropy of the density
never decrease. These requirements set $h^{0j}=0$ and
$h^{j0}=0$ respectively. Then we get
\begin{equation}
\frac{d}{dt} \rho =2\epsilon^{ijk} H^i \rho^j \sigma ^k
-h^{ij} \sigma ^i \rho^j
\end{equation}
Because the antisymmetric part of $h^{ij}$ can be absorbed into $H^i$, we
may assume that $h^{ij}$ is symmetric.
For $K^0-\overline{K^0}$ system, EHNS further assume that this new effect
does not change strangeness. So, $h^{1j}=0$ and
\begin{equation}
h=2\left( \begin{array}{ccc} 0 & 0 & 0 \\
0 & \alpha & \beta \\ 0 & \beta & \gamma \end{array}\right)
\end{equation}
where $\alpha,\beta,\gamma$ are the EHNS parameters and [10]
\begin{equation}
\alpha,\;\;\;\; \gamma >0, \;\;\;\;\;\;\;\;\; \alpha\gamma>\beta^2
\end{equation}

Notice that the neutrino oscillation among different flavors is very
similar to the strangeness oscillation in $K^0-\overline{K^0}$ system,
we think that it is reasonable to generalize the above assumption from
$K^0-\overline{K^0}$ system to neutrino system.

Here, three EHNS parameters are present. As we have mentioned before, all
the consideration from either theory or experiments by Ellis et al and
Peskin et al only give their upper limits. We have not yet known their
definite values - if this mechanism is a correct physical law. The central
purpose of this paper is intend to determine their lower limits constraint
by neutrino oscillation.\\

\noindent {\bf III. Neutrino Oscillation and the Present Situation
of Solar Neutrino Problem}

The neutrino weak eigenstates may not coincide with their mass eigenstates.
Because of the different time evolution, oscillation can occur[13-15].
For simplicity, we only consider the case of two flavors.
Suppose that the neutrinos mix through a vacuum mixing angle $\theta$, i.e,
\begin{equation}
\left (
\begin{array}{c}
 |\nu_e>\\
 |\nu_\mu>
\end{array}
\right )
=
\left (
\begin{array}{cc}
cos\theta & sin\theta \\
-sin\theta &cos\theta
\end{array}
\right )
\left (
\begin{array}{c}
 |\nu_1>\\
 |\nu_2>
\end{array}
\right )
\end{equation}        
where $|\nu_e>$, $|\nu_\mu>$ are the weak eigenstates and
$|\nu_1>$, $|\nu_2>$ the mass eigenstates with masses $m_1$ and $m_2$
respectively.
Because the two states evolve differently,
\begin{equation}
\left (
\begin{array}{c}
 |\nu_e(t)>\\
 |\nu_\mu(t)>
\end{array}
\right )
=
\left (
\begin{array}{cc}
cos\theta e^{-iE_1t }& sin\theta e^{-iE_2t }\\
-sin\theta e^{-iE_1t }&cos\theta e^{-iE_2t }
\end{array}
\right )
\left (
\begin{array}{c}
 |\nu_1>\\
 |\nu_2>
\end{array}
\right )
\end{equation}
As a result, a state start with $|\nu_e>$ may oscillate into $|\nu_\mu>$
with the probability
\begin{equation}
p(\nu_e\rightarrow\nu_\mu(t))=sin^2(2\theta)sin^2(\frac{1}{2}(E_2-E_1)t)
\end{equation}
and the probability for it to remain as itself is:
\begin{equation}
p(\nu_e\rightarrow\nu_e(t))=1-sin^2(2\theta)sin^2(\frac{1}{2}(E_2-E_1)t)
\end{equation}
Notice that the neutrino mass is very small, so their energy and momentum
are very close. Hence we can rewrite the survival probability as
following[21]:
\begin{equation}
p(\nu_e\rightarrow\nu_e(t))=1-sin^2(2\theta)sin^2(\frac{\delta m^2L}{4E})
\end{equation}
where $\delta m^2=m_{\nu_\mu}^2-m_{\nu_e}^2$.
Numerically, $\delta m^2L/(4E)=(1.266932\ldots)\delta m^2L/E$ , here
$\delta m^2$ is measured in $eV^2$,  $L$ in $m$ while $E$ in $MeV$
(or $L$ in $Km$ while $E$ in $GeV$). Because in all cases of this paper,
$p(\nu_e\rightarrow\nu_e(t))+p(\nu_e\rightarrow\nu_\mu(t))=1$, i.e., the
probability is conserved, so we only need to write down one of them.

Now, let's turn to the solar neutrino problem. This problem was first put
forward by the Homestake chlorine experiment around 1970. It was recognized
that the observed  capture rate was significantly smaller than the standard
solar model(SSM) prediction. After the middle of 1980s, some other experiment
groups began to collect datum. The recent results from the four solar
neutrino experiments[21-25] and the SSM predictions calculated by Bahcall and
Pinsonneault(B-P)[26] and by Turck-Chieze and Lopes(T-C-L)[27] are listed
in the following table.
\begin{table}[hp]
\caption{Experimental and Theoretical Datum of the Solar Neutrino}
\begin{center}
\begin{tabular}{l|c|c|r}
\hline\hline
{\sl Experiment} & {\sl Data} & {\sl B-P} & {\sl T-C-L} \\ \hline
Homestake & $2.55 \pm 0.17 \pm 0.18$ & $8.0 \pm 3.0$ & 6.4  \\
GALLEX    & $79 \pm 10 \pm 6$ & $131.5_{-17}^{+21}$ & 122.5  \\
SAGE      & $73_{-16-7}^{+18+5}$ & $131.5_{-17}^{+21}$ & 122.5 \\
Kamiokande & $2.89_{-0.21}^{+0.22} \pm {0.35}$ & $5.7 \pm 2.4$ & 4.4 \\
\hline\hline
\end{tabular}
\end{center}
\end{table}
Where, for Homestake, GALLEX and SAGE, the data are capture rates given in
solar neutrino units ($SNU. 1 SNU=10^{-36}$ capture per atom per second). For
Kamiokande, the datum is $^8B$ solar neutrino flux given in units of
$10^6 cm^{-2} s^{-1}$. The first errors are statistical and the second
errors are systematic. The errors associated with the B-P calculation are
"theoretical" $3$ standard deviations according to the authors.\\

\noindent{\bf IV. A Constraint on EHNS Parameters from
                  Solar Neutrino Problem}

Let's first review the effect of EHNS mechanism on neutrino oscillation.
For neutrino, its Hamiltonian is diagonalized in the basis of $|\nu_1>,
|\nu_2>$,
\begin{equation}
H=\left( \begin{array}{cc} E_1 & 0 \\
0 & E_2 \end{array}\right)
\end{equation}
so,
\begin{eqnarray}
H^1&=&H^2\;=\;0 \nonumber\\
H^3&=&(E_1-E_2)/2\; \approx\;-\delta m^2/(4E)
\end{eqnarray}
and Eq.(7) becames
\begin{equation}
\frac{d}{dt}\left( \begin{array}{c} \rho^0 \\ \rho^1
\\ \rho^2 \\ \rho^3 \end{array}\right)=2\left(
\begin{array}{cccc} 0 & 0 & 0 & 0
\\ 0 & 0 & \delta m^2/(4E) & 0
\\ 0 & -\delta m^2/(4E)  & -\alpha  & -\beta
\\0 & 0 & -\beta & -\gamma \end{array}\right)
\left( \begin{array}{c} \rho^0 \\ \rho^1
\\ \rho^2 \\ \rho^3 \end{array}\right)
\end{equation}
we only need to consider the case which the neutrino is originally
in the state $| \nu_e >$, from Eq.(10)
\begin{equation}
\rho_{\nu_e}=\left( \begin{array}{cc} cos^2 \theta & cos \theta sin \theta \\
cos \theta sin \theta & sin^2 \theta \end{array}\right)
\end{equation}
Then the initial condition is:
\begin{equation}
\rho(t=0)=\rho_{\nu_e}
\end{equation}
i.e:
\begin{equation}
\rho^0=\frac{1}{2}\;\;\;\;\;\;\;
\rho^1=\frac{1}{2}sin(2\theta)\;\;\;\;\;\;\;
\rho^2=0\;\;\;\;\;\;\;
\rho^3=\frac{1}{2}cos(2\theta)
\end{equation}
The survival probability of $|\nu_e>$ is
\begin{equation}
p(\nu_e\rightarrow\nu_e)=Tr\lbrack\rho(t)\;\;\rho_{\nu_e}\rbrack
\end{equation}
where $\rho(t)$ is given by Eq.(17) and Eq.(2).

Solving Eq.(17) analytically and exactly is very tedious. In this work,
we suppose that $\delta m^2/(4 E)$ is at least one magnitude order larger
than $\alpha, \;\; \beta, \;\; \gamma $. In fact, this can always be
satisfied for solar neutrino problem. For example, for the small- and
the large-mixing solution, $\delta m^2/(4 E)$ is about the order of
$10^{-21} \;\; GeV$[21] if we take the magnitude order of the solar neutrino
energy as $1\; MeV$, but the upper limit of $Max[\alpha,\; \beta, \;
\gamma]$ given by J.Ellis et al.[20] for the $K^0-\overline{K^0}$ system
is $4 \times 10^{-17} \;\; GeV$, when transforming it into the neutrino
system, as discussed in [19], its magnitude is about $10^{-22} \;\; GeV$.
This is just one magnitude order less than $\delta m^2/(4 E)$.

When approxmite to the first order, we can write down the neutrino survival
probability analytically as
\begin{equation}
p(\nu_e\rightarrow\nu_e) \simeq \frac{1}{2}\;+
\;\frac{1}{2}\;cos^2(2 \theta)\;e^{-2 \gamma L}\;+
\;\frac{1}{2}\;sin^2(2 \theta)\;cos(\frac{\delta m^2}{2 E}L) \;e^{-\alpha L}
\end{equation}

From Eq.(22), It is obviously that as $L \rightarrow \infty ,
\;\; p(\nu_e\rightarrow\nu_e(t)) \rightarrow \frac{1}{2} $.
This is the most important result of EHNS mechanism on neutrino oscillation
in the case of two generations.

To generalize the case of two generations to the case of three generations,
we can begin with Eq.(1), expand the Hamilitonian $H$ and the density
matrix $\rho$ by using the generators of $SU(3)$, then eight EHNS parameters
which have no any experiental restriction will be presence, so, such
generation is meaningless.

Here, we further suppose that the direct oscillation between $\nu_e$ and
$\nu_\tau$ can be neglect, we only consider the case that the oscillation
between $\nu_e$ and $\nu_\tau$ takes the following way: first, $\nu_e$
oscillates into $\nu_\mu$, then, due to the oscillation between $\nu_\mu$
and $\nu_\tau$, it oscillates into $\nu_\tau$. If the system starts with
$|\nu_e>$, then the oscillation looks like a waterfall. After making such
supposition, we can deal with the neutrino oscillation by including ENHS
mechanism in the case of two generations. Due to the discussion of the effect
of EHNS mechanism on the oscillation between $\nu_\mu$ and $\nu_\tau$ is the
same as that between $\nu_e$ and $\nu_\mu$, we will not repeat it here. But
it is evidently that because of the effect of EHNS mechanism,
$$
L \rightarrow \infty , \;\; p(\nu_e\rightarrow\nu_e(t))
\rightarrow \frac{1}{3}
$$
when we consider the case of three generations.

Now, return to table 1, it is easily to see that the ratios of experimental
central values to theoretical values are between
$$
0.32\;\; \sim \;\; 0.6.
$$On the
other hand, if the EHNS effect on neutrino oscillation has not yet developed
sufficiently, i.e., the probability decay has not yet developed
sufficiently, then the survival probability happens to local in this domain.
Hence we can get the experimental constraint on the EHNS parameters.

A detail calculation of the dependence of the survival probability
$p(\nu_e\rightarrow\nu_e)$ on the time $T$ or distance $L$ is difficult,
as a estimation, we have
\begin{equation}
\frac{1}{2}\;+\;\frac{1}{2}\;cos^2(2 \theta)\;e^{
-2 \gamma L}\;+\;\frac{1}{2}\;sin^2(2 \theta)\;
cos(\frac{\delta m^2}{2 E}L)\; e^{-\alpha L}
\leq 0.6
\end{equation}
where $\theta$ is the vacuum mixing angle of neutrino and
$$
L\simeq 7.58\times 10^{26} GeV^{-1}
$$
is the distance
from the sun to the earth in natural units.

Eq.(23) is the most important result in this work, it is just the
constraint on EHNS parameters that we want to find, from which we can
extract some concrete conclusion.

From Eq.(23), we achieve,
\begin{equation}
sin^2(2 \theta)\;\{\;\left [1\;-\;2 sin^2(\frac{\delta m^2}{4 E} L)
\right ]\;e^{- \alpha L}\;-\;e^{-2 \gamma L}\;\}
+\; e^{-2 \gamma L}
\leq 0.2
\end{equation}
\begin{equation}
cos^2(2 \theta)\;\{\;e^{-2 \gamma L}\;-\;e^{-\alpha L}
\left [ 1+2 tg^2(2 \theta) sin^2(\frac{\delta m^2}{4 E}L) \right ]
 \;\}+\; e^{-\alpha L}
\leq 0.2
\end{equation}
Hence, we obtain
$$
e^{-2 \gamma L}\; \leq \; 0.2 \;\;\;\;\;
if\;\;\;\; \alpha < 2\; \gamma +
\frac{1}{L}Log\left [1\;-\;2 sin^2(\frac{\delta m^2}{4 E} L)
\right ]
$$
and
$$
e^{-\alpha L}\; \leq \; 0.2 \;\;\;\;\;
if\;\;\;\; \alpha > 2\; \gamma+
\frac{1}{L}Log\left [ 1+
2 tg^2(2 \theta) sin^2(\frac{\delta m^2}{4 E}L) \right ].
$$
Notice that
$Log[1\;-\;2 sin^2(\frac{\delta m^2}{4 E} L)]<0$ and
$Log[ 1+2 tg^2(2 \theta) sin^2(\frac{\delta m^2}{4 E}L)]>0$, then we arrive
$$
\gamma\; \geq \; 1.06\times 10^{-27}\;GeV
\;\;\;\;\; if\;\;\;\; \alpha < 2\; \gamma
$$
and
$$
\alpha\; \geq \; 2.12\times 10^{-27}\; GeV
\;\;\;\;\; if\;\;\;\; \alpha > 2\; \gamma
$$

But according to EHNS and Huet and Peskin, when transfering these parameters
produced in neutrino system into $K^0-\overline{K^0}$ system, a certain
ratio should be considered. As discussed in [19], we will multiply
them by a factor of $(\frac{500 MeV}{E_\nu (in MeV)})^2$.
Where as a estimate, we take $E_\nu$ as the average energy of the solar
neutrino, then $E_\nu\sim 0.6 MeV$[21]. Finally, we get
\begin{equation}
\gamma\; \geq \; 7.4\times 10^{-22}\;GeV
\;\;\;\;\; if\;\;\;\; \alpha < 2\; \gamma
\end{equation}
and
\begin{equation}
\alpha\; \geq \; 1.5\times 10^{-21}\; GeV
\;\;\;\;\; if\;\;\;\; \alpha > 2\; \gamma.
\end{equation}
This is consistent with their upper limits[20], $\alpha\leq 4\times 10^{-17}
GeV$ and $\gamma \leq 7\times 10^{-21} GeV$. \\

\noindent{\bf V. Conclusion and discussions }

In conclusion, we have derived a lower limit constraint on the EHNS
parameters from the solar neutrino problem by supposing that the defect
of the solar neutrino is only due to the EHNS mechanism. We find
$\gamma\; \geq \; 7.4\times 10^{-22}\;GeV
$
if
$
\alpha < 2\; \gamma
$
and
$
\alpha\; \geq \; 1.5\times 10^{-21}\; GeV
$
if
$
\alpha > 2\; \gamma.
$ This constraint is coincide with the latest result given
by Ellis et al.[20] and can be detected in the future experiments.

Here, the uncertainty arise in the amplification factor
$(\frac{500 MeV}{E_\nu (in MeV)})^2$. Because the solar neutrino
energy spectrum has a certain width, so when we do not take its
average value as the input parameter of the amplification factor,
departure will be present, and the lower limits will
be risen for the sector of small energy. But, notice that the present
experiments have their threshold, even if we take $E_\nu$ as $0.2\;MeV$,
the lower limits constraint determinated above is still consistent with
Ellis et al.[20] and Huet and Peskin[10].

In this work, we have supposed that the solar neutrino problem
is only due to the EHNS mechanism. We have not considered the MSW
effect yet. The detail discussion on how about the correction due to
the MSW effect will be reported in the future.\\

\noindent Acknowledgment: One of the authors (Y.Liu) would like to
thank prof. Xueqian Li for leading him into this research field and
prof. Dongsheng Du for his encourgement.


\begin{thebibliography}{99}
\bibitem{1} S.W.Hawking, Nature 248(1974)30.
\bibitem{2} S.W.Hawking, Commun.Math.Phys. 43(1975)199.
\bibitem{3} T.Banks, M.E.Peskin and L.Susskind, Nucl.Phys.B 244(1984)125.
\bibitem{4} D.N.Page, Gen.Rel.Grav. 14(1982)1; Phys.Rev.Lett. 44(1980)301.
\bibitem{5} L.Alvarez-Gaume and C.Gomez, Commun.Math.Phys. 89(1983)235.
\bibitem{6} J.Ellis, J.S.Hagelin, D.V.Nanopoulos and M.Srednicki, Nucl.Phys.B
 241(1984)381.
\bibitem{7} B.Reznik, Phys.Rev.Lett. 76(1996)119.
\bibitem{8} J.Ellis, N.E.Mavromatos and D.V.Nanopoulos, Phys Lett.B
 293(1992)142.
\bibitem{9} CPLEAR Collaboration and J.Ellis et al. Phys.Lett.B
364(1995)239
\bibitem{10} P.Huet and E.Peskin, Nucl.Phys.B 434(1995)3. and the refrences
therein.
\bibitem{11} J.Ellis, N.E.Mavromatos and D.V.Nanopoullos,
Phys.Lett.B 293(1992)37.
\bibitem{12} J.Ellis, N.E.Mavromatos, E.Winstanley and D.V.Nanopoulos,
Mod.Phys.Lett.A 12(1997)243.
\bibitem{13} L.Wolfenstein, Phys.Rev.D 17(1978)2369; 20(1979)2634.
\bibitem{14} S.P.Mikheyev and A.Yu.Smirnov, Yad.Fiz. 42(1985)1441(
Sov.J.Nucl.Phys.42(1985)913); Nuov.Cimento.C 17(1986)9.
\bibitem{15} V.Barger, K.Whisnaut, S.Pakvasa and R.J.N.Phillips, Phys.Rev.D
22(1980)2718.
\bibitem{16} D.N.Spergel and W.H.Press, Astrophys.J. 294(1985)663;
W.H.Press and D.N.Spergel, AstroPhys.J. 296(1985)679.
\bibitem{17} J.Faulkner and R.Gilliland, Astrophys.J. 299(1985)994.
\bibitem{18} J.Pulido, Phys.Rept. 211(1992)167; G.Gelmini and E.Roulet,
Rep.Prog.Phys. 58(1995)1207.
\bibitem{19} Y. Liu, L.Z.Hu and M-L Ge, Effect of Violation of Quantum
Mechanics on Neutrino Oscillation. In press in Phys.Rev.D.
\bibitem{20} J.Ellis, J.L.Lopez, N.E.Mavromatos and Nanopoulos, Phys.Rev.D
53(1996)3846.
\bibitem{21} Particle Data Group, Phys.Rev.D 54(1996)1.
\bibitem{22} B.T.Cleveland et al., Nucl.Phys.(Proc.Supp.)B 38(1995)47.
\bibitem{23} P.Anselmann et al., Phys.Lett.B327(1994)377.
\bibitem{24} J.N.Abdurashitov et al., Nucl.Phys.(Proc.Supp.)B 38(1995)60.
\bibitem{25} Y.Suzuki, Nucl.Phys.(Proc.Supp.)B 38(1995)54.
\bibitem{26} J.N.Bahcall and M.H.Pinsonneault, Rev.Mod.Phys. 64(1992)885.
\bibitem{27} S.Turck-Chieze and I.Lopes, Astrophys.J. 408(1993)347.

\end{thebibliography}
\end{document}